\documentclass{elsart}

 \usepackage{graphicx}

\usepackage{amssymb}

\begin{document}

\begin{frontmatter}



\title{Relativistic effects of rotation in $\gamma$-ray pulsars - Invited Review}


\author{Osmanov Z.N.} 

\address{School of Physics, Free University of Tbilisi, 0183, Tbilisi, Georgia}

\address{E. Kharadze Georgian National Astrophysical Observatory, Abastumani 0301, Georgia}

\begin{abstract}
In this paper we consider relativistic effects of rotation in the magnetospheres of $\gamma$-ray pulsars. The paper reviews the progress achieved in this field during the last three decades. For this purpose we examine direct centrifugal acceleration of particles and corresponding limiting factors: constraints due to the curvature radiation and the inverse Compton scattering of electrons against soft photons. Based on the obtained results generation of parametrically excited Langmuir waves and the corresponding Landau-Langmuir-centrifugal drive is studied.

\end{abstract}


\end{frontmatter}

\section{Introduction}

One of the fundamental problems of $\gamma$-ray pulsar astrophysics is the generation of VHE emission. In this context it is worth noting that last decade has been very fruitful in detecting VHE $\gamma$-rays of pulsars \cite{list1,list2,list3,list4,list5}. On the other hand, it is clear that generation of VHE emission requires existence of relativistic particles and the process of their acceleration is still under debate. 

Generally speaking, it is observationally evident that the pulsars are rotating neutron stars with increasing period of rotation, $P$, characterized by the slow down rate, $\dot{P}\equiv dP/dt>0$. The corresponding slow down power $\dot{W} = I\Omega|\dot{\Omega}|$, then becomes 
\begin{equation}
\label{power} 
\dot{W}\simeq 4.7\times 10^{31}\times P^{-3}\times\frac{\dot{P}}{10^{-15}}\times\frac{M_{ns}}{1.5M_{\odot}}\; ergs\; s^{-1},
\end{equation} 
where the slow down rate is normalized by the typical value for normal period pulsars (with period of the order of one second) and we have taken into account that the moment of inertia of a sphere is expressed as $I = 2M_{ns}R_{\star}^2/5$, $M_{ns}$ denotes the neutron star's mass, $R_{\star}\simeq 10^6$ cm is its radius and $M_{\odot}\simeq 2\times 10^{33}$ g is the solar mass. It is clear from the aforementioned expression that slowing down of pulsar's rotation provides enormous power, which might be even higher for millisecond pulsars. Therefore there is no question about a source of energy, but the problem concerns a particular mechanism which might be responsible for acceleration of particles up to relativistic energies. 

In the framework of the so-called polar-cap models by means of the strong electric field parallel to magnetic field lines (parallel electric field) the particles are uprooted from the surface of the neutron star \cite{ruds}, generation of electric fields in turn, is a direct result of neutron star's rotation. In the vicinity of the neutron stars the electrons might undergo further acceleration provided either by magnetic field line curvature \cite{arons}, or space-charge-limited flow \cite{michel}, or effects of inertial frame \cite{musl}. On the other hand, generation of parallel electric field might be strongly suppressed either by means of the field curvature \cite{DH} or the IC scattering of electrons against photons \cite{dermer}. Therefore, to explain the origin of VHE particles outer gap models became actual. In the framework of this approach the electrons gain energy in the outer magnetosphere in the so called vacuum gap \cite{cheng,hirotani}, but as it turned out, in these scenarios the gap size is not enough to explain the observed VHE emission from pulsars. Some attempts to enlarge the gap size \cite{AS,US,musl} although resulted in a more efficient acceleration process, but still could not solve the problem. By considering the mechanism of energization along the magnetic field lines in the force-free pulsar magnetospheres has been studied in detail by Contopulous et al. (see \cite{contop}) where the authors have pointed out that acceleration is significantly suppressed by the force-free topology of the magnetic field. The acceleration process of a two component electron-positron plasma in a monopole configuration near the LC (a hypothetical zone, where the linear velocity of rotation exactly equals the speed of light) area has been examined in \cite{beskin} and it has been shown that the maximum Lorentz factors of particles is of the order of $\sigma\equiv B^2/(8\pi nm_ec^2)$, where $B$ is the magnetic field, $n$ is the particle number density, $m_e$ is the electron's mass and $c$ is the speed of light.

\begin{figure}
 \centering
  \resizebox{5cm}{!}{\includegraphics[angle=0]{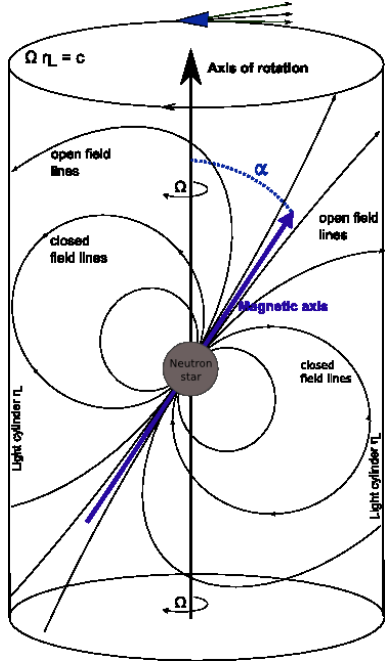}}
  \caption{Sketch of the magnetospheric Pulsar model from \cite{OR2}.
  The centrifugally accelerated co-rotating particles in the nearby zone of the LC area emit in a relatively narrow cone along the tangential direction.} \label{fig1}
\end{figure}

A rather different mechanism of acceleration which might take place in the LC zone has been proposed by Gold \cite{gold1,gold2}, who suggested that extremely strong magnetic field forces particles to slide along the field lines, which are trajectories of particles in the rotating frame of reference. This is an effective picture: the electrons gyrate in magnetic field and a trajectory averaged by time for long time-scales is a certain trajectory, which in a very strong magnetic field approximation is a line, along which the particles slide. In particular, the magnetic field in the pulsar's magnetosphere is assumed to be dipolar $B\simeq B_{st}\times (r_s/r)^3$, where $B_{st}\simeq 3.2\times 10^{19}\sqrt{P\dot{P}}$ Gauss is the magnetic induction on the neutron star's surface \cite{MT} and $r_s\simeq 10$ km is the pulsar's radius. Then, for electrons with relativistic factors of the order of $10^5$, the gyro-radius in the LC area (a hypothetical zone where the linear velocity of rigid rotation exactly equals the speed of light) for normal period pulsars is of the order of $0.0016$ cm, which is by many orders of magnitude than the corresponding length-scale of the LC radius: $10^9$ cm. Therefore, the electrons are  frozen into the magnetic field and follow the co-rotating field lines. The corresponding condition of the frozen-in state ${\bf E}+\frac{1}{c}{\bf v}\times{\bf B}= 0$ explains the drift of particles in the direction of rotation as follows: by means of the radial magnetic field and the generated electric field perpendicular to the equatorial plane the  ${\bf E}\times{\bf B}/c^2$ drift occurs, which actually is the co-rotation velocity, ${\bf v}$, of the particle.

On the other hand, it is clear that the total velocity cannot be greater than the speed of light and therefore, because of the co-rotation of the pulsar's magnetosphere \cite{ruderman}, by reaching the LC surface (if the rigid rotation is preserved) the longitudinal velocity will tend to zero and electrons will have only the azimuthal component, leading to the emission along the tangential direction Fig. \ref{fig1}. Here $\Omega$ denotes the angular velocity of neutron star's rotation, $r_{_L}$ is the radius of the LC and $\alpha$ is the angle between the dipole magnetic moment and the axis of rotation.

In this review paper we consider the centrifugally generated relativistic electrons as for normal period pulsars as well as for millisecond ones and present the progress achieved in this field during almost the last three decades. The paper is organized in the following way: Sec. 2 gives a detailed description of different phenomena which take place due to the centrifugal effects. We apply the model of centrifugal acceleration to the well known Crab and Vela pulsars and normal period pulsars and in Sec. 3 we outline the obtained results.

\section{Results}

In this section we consider effects of rotation of the pulsar's magnetosphere in the LC area and study several problems: Centrifugal acceleration of particles; Rotationally driven Langmuir waves; Landau Langmuir centrifugal drive and reconstruction of pulsars' magnetospheres by means of the centrifugal effects.

\subsection{Centrifugal acceleration}

As we have already mentioned in the introduction, the longitudinal velocity component should decrease in the nearby zone of the LC. Deceleration of the radial velocity and thus the centrifugal force reversal has been investigated in the Schwarzschild black-hole metrics \cite{abramow}. The corresponding problem but in a special relativistic case has been studied in \cite{MR}, where authors have considered a gedanken experiment: a bead freely sliding on a rotating wire and studied particle's dynamics. It has been demonstrated that effect of "relativistic mass" leads to the similar phenomenon: if the rigid rotation is preserved, the radial velocity gradually decelerates and vanishes on the LC zone. For studying the problem of acceleration one can perform calculations in the co-rotating reference frame. then, if one introduces an interval for a straight field line inclined by the angle $\alpha$ with respect to the rotating axis \cite{MR,incr}
\begin{equation}
\label{ds} 
ds^2 = -c^2dt^2\left(1-\frac{\Omega_{ef}^2r^2}{c^2}\right)+dr^2,
\end{equation} 
where $\Omega_{ef}\equiv\Omega\sin\alpha$ is the effective angular velocity of rotation for a field line inclined by the angle $\alpha$ with respect to the rotation axis.

The particles moving along the co-rotating trajectories are characterised by the following Lagrangian
\begin{equation}
\label{lag} L =
\frac{1}{2}G_{\mu\nu}\frac{dx^{\mu}}{d\tau}\frac{dx^{\nu}}{d\tau},
\end{equation}
and the corresponding equation of motion

\begin{figure}
 \centering
  \resizebox{10cm}{!}{\includegraphics[angle=0]{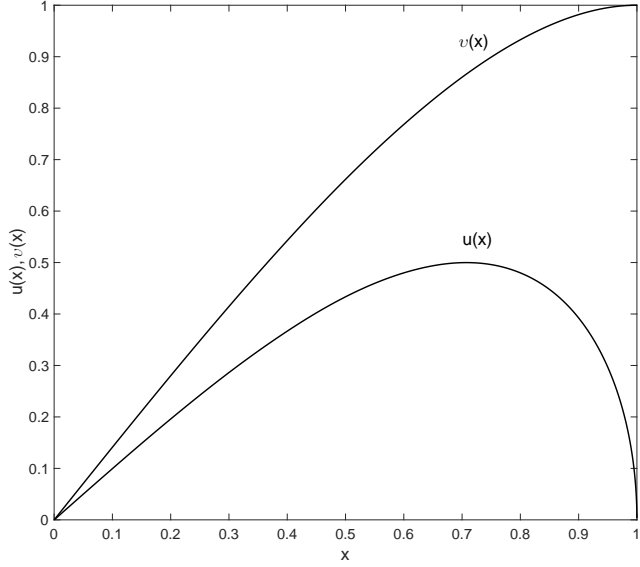}}
  \caption{Here we plot  $u$ and $\upsilon$ versus dimensionless radial distance, $x$. The set of parameters is $x_0 = 0.001$ and $u_0 = 0$.} \label{fig2}
\end{figure}

\begin{equation}
\label{equat} \frac{\partial L}{\partial x^{\mu}} =
\frac{d}{d\tau}\left(\frac{\partial L}{\partial
\dot{x}^{\mu}}\right)
\end{equation}
where $G_{00} = -(1-\Omega_{ef}^2r^2/c^2)$, $G_{11} = 1$, $\mu,\nu = \{0,1\}$, $x^0 = ct$, $x^1 = r$.

Considering the radial component of the equation of motion one can show that the corresponding equation in the dimensionless form is given by
\begin{equation}
\label{dv} 
\frac{d^2x}{d\tau^2} = \frac{x}{1-x^2}\left[1-x^2-2\left(\frac{dx}{d\tau}\right)^2\right],
\end{equation} 
where $x\equiv r/r_{_L}$, $r$ is the radial coordinate, $r_{_L}=c/\Omega$ denotes the LC radius, $\tau\equiv t\Omega$ is the dimensionless time, $dx/d\tau=\upsilon_r/c\equiv u$ and $\upsilon_r$ denotes the radial velocity. In Fig. 2 we show the behaviour of  radial velocity, $u(x)$ and total velocity, $\upsilon (x)\equiv\sqrt{x+u(x)^2}$ in the dimensionless forms for the following initial values $x_0 = 0.001$ and $u_0 = 0$ (note that the linear velocity of rotation in the mentioned dimensionless form coincides with the radial coordinate). As it is clear from the plots, initially the radial velocity increases, but in due course of time, by reaching the LC ($x = 1$) it starts decelerating and the total velocity of the particle tends to the speed of light ($\upsilon\rightarrow 1$). By considering a plasma flow in the co-rotating pulsar's magnetosphere it has been shown that in the limit of strong magnetic field the system of fluid equations reduce to a single particle approach described by Eq. (\ref{dv}). 

Eq. (\ref{lag}) for $\mu = 0$ leads to energy - constant of motion
\begin{equation}
\label{ener} 
E = \gamma mc^2\left(1-\frac{\Omega r}{c^2}\upsilon_{\varphi}\right),
\end{equation} 
where $m$ is the particle's mass, $\upsilon_{\varphi}$ is the azimuthal velocity component and $\gamma = \left(1-\upsilon_r^2/c^2-\upsilon_{\varphi}^2/c^2\right)^{-1/2}$ is the Lorentz factor of the particle. If one imposes the condition of constancy the radial behaviour of the relativistic factor writes as \cite{FrankR,OR2}
\begin{equation}
\label{gamma} 
\gamma (r)  = \gamma_0\frac{1-r_0^2/r_{_L}^2}{1-r^2/r_{_L}^2},
\end{equation} 
where $\gamma_0$ and $r_0$ denote initial values of the Lorentz factor and radial coordinate respectively. Eq. (\ref{gamma}) confirms that particle's energy asymptotically increases close to the LC.

From the observations it is clear that particles originating in the pulsar magnetospheres, leave these regions and dynamically become force-free. This means that the trajectory of the particle in the laboratory frame of reference is a straight line. Since the particles are in the frozen-in condition and follow the magnetic field lines, the shape of the latter in the co-rotating frame of reference should be given by the Archimedes' spiral, lagging behind the rotation \cite{RDO}. On the other hand, some of the $\gamma$-ray pulsars exhibit jet-like structures \cite{jet}, therefore one should introduce 3D configuration of magnetic field lines co-rotating with the neutron star. It is worth noting that the field lines will become more and more twisted in a narrow region of the LC. This particular problem has been considered in detail in \cite{ODM}. It was shown that in the nearby zone of the LC the curvature drift of plasma particles driving perpendicular to the equatorial plane leads to generation of a toroidal component. This in turn, changes the shape of the field lines into a configuration of the Archimedes' spiral. For this purpose, in the co-rotating frame of reference we examine the field line configuration 
\begin{equation}
\label{line} \varphi = \varphi (r),~~ z = f(r).
\end{equation}
Imposing the aforementioned prescribed trajectory on the the space-time metric
\begin{equation}
\label{metr} ds^2 = - dt^2 + r^2 d\phi^2 + dr^2 + dz^2,
\end{equation}
will reduce it to the form
\begin{equation}
\label{met} ds^2 = G_{00}dt^2+2G_{01}dtdr+G_{11}dr^2,
\end{equation}
with
\begin{equation}
\label{gab1} G_{\alpha\beta} = \left(\begin{array}{ccc}
-1+\omega^2r^2, \;\;\; & \omega\varphi^{\prime}\rho^2 \\
\omega\varphi^{\prime}r^2, \;\;\;& 1+\varphi^{\prime 2}r^2 +
\acute{f}^2 \\
\end{array}\right),
\end{equation}
where $\varphi^{\prime} = d\varphi/dr$, $f^{\prime} = df/dr$, $c = 1$. Then, in the same manner as before, one can derive an expression for the energy for the non-diagonal metrics 
\begin{equation}
\label{energ} E = -\gamma\left(G_{00}+G_{01}v\right) = const,
\end{equation}
where $v\equiv dr/dt$ is the radial velocity and
\begin{equation}
\label{gam}
d \tau/dt \equiv u^t  \equiv\gamma = \left(-G_{00}-2G_{01}v-G_{11}v^2\right)^{-1/2}
\end{equation}
is an expression for the relativistic factor in the general case - $3D$ curved trajectories. One can explicitly solve Eqs. (\ref{energ},\ref{gam}) for the radial velocity
\begin{equation}
\label{vv} v={\frac{\sqrt{G_{00}+E^2}}{(G_{01}^2+E^2G_{11})}}
{\left[-G_{01}\sqrt{G_{00}+E^2} \pm E \sqrt{G_{01}^2-G_{00}G_{11}},
\right]},
\end{equation}
where different signs correspond to different initial conditions.

\begin{figure}
 \centering
  \resizebox{10cm}{!}{\includegraphics[angle=0]{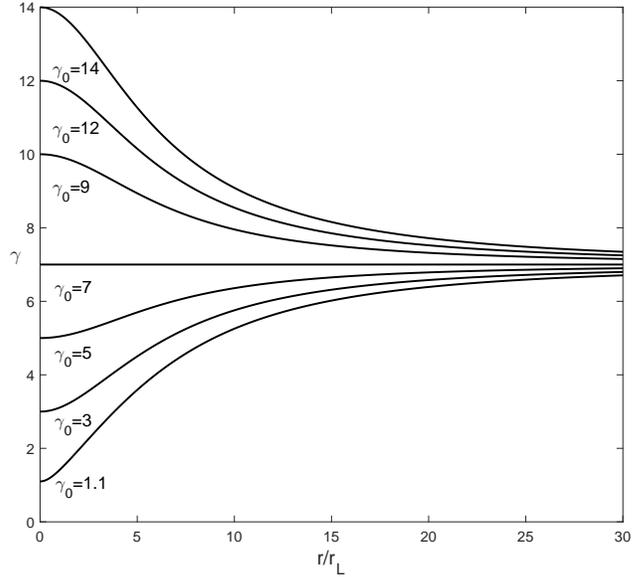}}
  \caption{The dependence of the Lorentz factor on the radial
coordinate for different initial relativistic factors $\gamma_0$.The calculations are done for an Archimedean spiral with the critical Lorentz factor $\gamma_{cr} = 7$. All particles are launched from $r_0 = 0$. As it is evident from the plot, the curve $\gamma_{cr}$ attracts all other curves. The figure is taken from \cite{OR2}.} \label{fig3}
\end{figure}

From the Fig. 2 it is clear that a particle sliding along a rectilinear co-rotating field lines never cross the LC zone. The reason follows from a nature of rigid rotation of straight field lines. On the other hand, as we have already mentioned the particles originating in the pulsars' magnetospheres might leave this region becoming dynamically free. This in turn, leads to a specific shape of the field lines in the local frame of reference - Archimede's spiral \cite{jet}. 

As a first example we consider a curved field lines having the shape of the Archimedean spiral located in the equatorial plane 
\begin{equation}
\label{case1} 
\varphi = ar,\; z = 0.
\end{equation}
For a spiral which lags behind the rotation ($a<0$) one can find a critical velocity when there is no "reaction force" acting from the curved trajectory. In this case the angular velocity of the particle in the laboratory frame, $\Omega_l = \Omega+a\upsilon$,  equals zero and therefore the corresponding critical radial velocity is defined as $\upsilon_{cr} = -\Omega/a$.

In Fig. 3 we plot the behaviour of the Lorentz factor versus the radial
coordinate for different initial relativistic factors $\gamma_0$.The calculations are performed for an Archimedean spiral with the critical Lorentz factor $\gamma_{cr} = 7$. All particles are launched from $r_0 = 0$. It is clear from the figure that if the initial value of $\gamma$ equals $7$ the particle does not experience the reaction force and therefore it stays constant. Unlike this case, if the initial values are different from the critical value of the relativistic factor, it asymptotically tends to $\gamma_{cr}$. From the plots is evident that particles go beyond the LC zone. The reason is following: since in due course of time the Lorentz factor tends to its critical value, the effective angular velocity, $\Omega_l$, tends to zero and consequently the corresponding LC radius "inflates" and the particle always stays inside $c/\Omega_l$ but outside the distance, $c/\Omega$.

A second example we would like to consider is the field lines located in a common plane with the rotation axis: 
\begin{equation}
\label{case2} 
\varphi = const,\; z = f(r).
\end{equation}
After imposing this condition, from Eq. (\ref{vv}) one can derive a longitudinal velocity (along a field line) as follows 
\begin{equation}
\label{vpar} v_{_{\parallel}} =
\left[\left(1-\omega^2r^2\right)\left(1-\frac{1-\omega^2r^2}{E^2}\right)\right]^{\frac{1}{2}}.
\end{equation}
where
\begin{equation}
\label{en} E = \gamma_{0}\left(1-\omega^2r_{0}^2\right),
\end{equation}
and $\gamma_0$ and $r_{0}$ are initial values quantities. Centrifugal mechanism of acceleration can explain the Crab pulsar's jet-like wind's velocity. From observations is evident that jets are collimated flows. Therefore, one can model them as the particles moving along the field lines with the following asymptotic structure: $r\rightarrow R$ and $z\rightarrow\infty$ ($R<r_{_L}$). The aforementioned velocity in the unit of the speed of light has its maximum value, which can be straightforwardly obtained from Eq. (\ref{vpar}) (see for details \cite{jet}) $v_{_{\parallel}}^{max} = 0.5$. On the other hand, the velocity averaged across the jet should be less than the possible maximum value, which is known from observations: $\upsilon_{Jet}\simeq 0.4$ \cite{jet}.

As a last example it is natural to consider a field configuration given by
\begin{equation}
\label{case3} 
\varphi = ar,\; z = f(r)
\end{equation}
which, on the one hand is characterised by a $3D$ geometry and one the other hand has a property of the Archimedean spiral.

After straightforward mathematical manipulations From Eq. (\ref{vv}) one can derive an asymptotic behaviour of the radial velocity 
\begin{equation}
\label{vas} v\rightarrow
-\frac{\omega}{a}+\frac{1}{r^2\omega a}
\left(E^2-1+\frac{E\sqrt{a^2-\omega^2\left(1+\acute{f}^2\right)}}{a}\right).
\end{equation}
It is evident from this expression that the effective angular velocity $\Omega_{l} \propto 1/r^2$ is a continuously decreasing function of the distance and asymptotically tends to zero, implying that the particle becomes dynamically force-free. 
\begin{figure}
  \resizebox{\hsize}{!}{\includegraphics[angle=0]{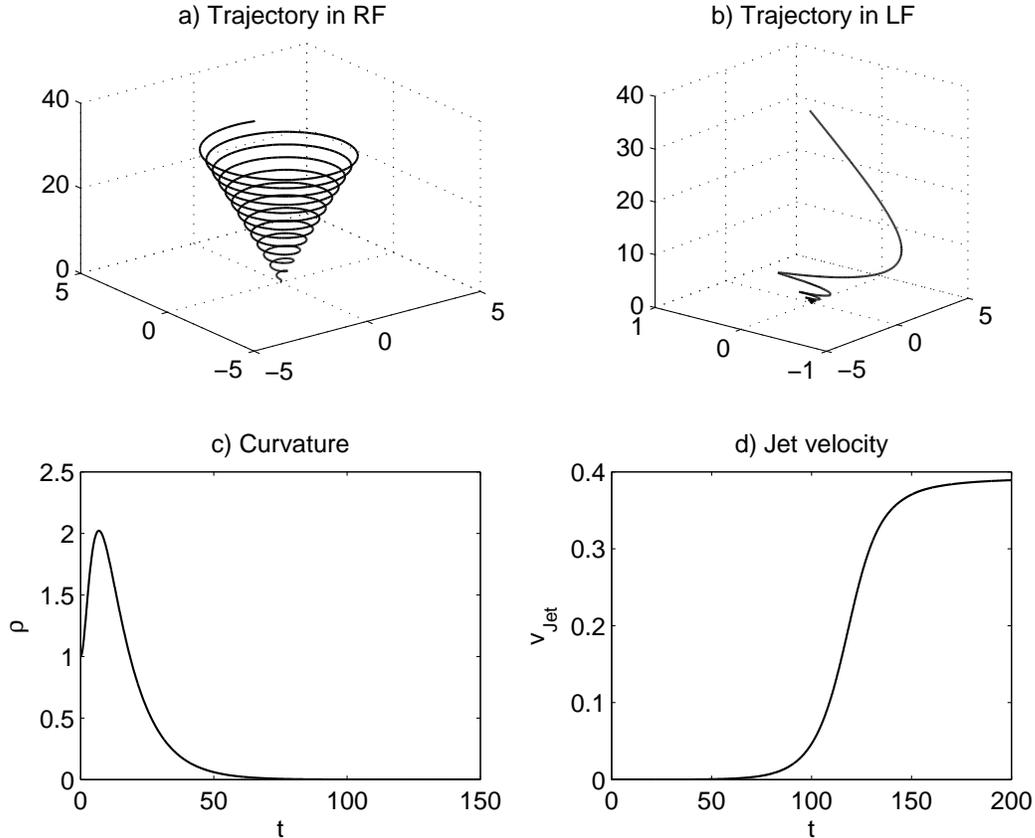}}
  \caption{The figure is from \cite{jet}. On the top panel we show the particle trajectories in
  (a) the rotational frame of reference and (b) the laboratory
  frame of reference respectively. On graphs (c) and (d) the behaviour of the curvature
  ($\rho\equiv 1/R_c$, where $R_c$ is the curvature radius of
  magnetic field lines) and jet velocity is shown respectively.
  The set of parameters is: $a = -28$,
  $f^{\prime}= 11$, $v_0 = 0.01$ and $r_0 = 0$.}\label{fig4}
\end{figure}

On Fig. 4 we show the trajectories of particles in (a) the rotational frame (RF) and (b) the laboratory frame (LF) respectively. we also show (c) the behaviour of the curvature (normalized by the initial value) of the trajectory in the LF and (d) the jet velocity, thus the velocity along the $z$ axis. The set of parameters is:  $a = -28$, $f^{\prime} = 11$, $v_0 = 0.01$ and $r_0 = 0$. It is clear that despite the fact that the particle's trajectory in the RF is given by a regularly shaped 3D-Archimedean spiral, in the LF the trajectory is more complex (see (b)). Initially the trajectory is more curved (see (c)) but in due course of time the curvature vanishes and as a result the trajectory in the LF becomes more rectilinear (see (b)). This means that dynamics tends to the force-free regime. Indeed, from Fig. 4 (d) one can see that the $z$ component of the velocity initially increases but asymptotically saturates indicating the transition to the zero force regime.

Since the magnetospheres of pulsars are full of photon field, the similar problems have been studied for a situation when a co-rotating system is imbedded in the isotropic photon field leading to a certain friction force influencing dynamics of particle acceleration. Likewise the previous cases it is shown that when the trajectories are presented by the Archimedean spiral with arms lagging behind the rotation, the particles reach force-free regime whatever the value of the friction force is. An interesting feature has been found for the Archimedean spiral with arms oriented in opposite direction. In this case  the particles do not tend to the force free regime, but instead reach a certain equilibrium location. The similar results have been obtained for 3D configurations of the field lines (for details a reader is referred to \cite{bakh1,bakh2}). 

\subsubsection{Co-rotation constraints}

It is natural to suppose that for almost straight field lines rigid rotation cannot be maintained up to the LC surface, therefore in order not to violate the relativity principle the pulsar's magnetosphere must reconstruct in a proper way. In particular, the rigid rotation is maintained by means of the strong magnetic field. Therefore, the co-rotation condition is satisfied if the magnetic field energy density, $B^2/8\pi$, exceeds the plasma energy density, $\gamma n_{_{GJ}}Mm_ec^2$, where $B$ denotes the magnetic field, $n_{_{GJ}} = {\bf \Omega B}/(2\pi e c)$ is the Goldreich-Julian number density of electrons in the pulsar's magnetosphere \cite{GJ}, $e$ is the electron's charge and $M$ is the multiplicity factor of particles. In this context one has to note that in the pulsar models it is assumed that primary particles, accelerated by the parallel electric field radiate due to the curvature radiation. The interaction of the emitted photons with the strong magnetic field will inevitably lead to the very efficient pair creation process ($\gamma+{\bf B}\rightarrow e^++e^-+{\bf B}$), newly generated secondaries repeat the same cascading mechanism until the plasma particles screen the electric field \cite{sturrock,tademaru}.

By taking the magnetic field strength and pulsar's radius $r_s\simeq 10$ km, into account combining with the aforementioned condition and Eq. (\ref{gamma}) one can show that for $r_0<<r$ the maximum attainable relativistic factor of electrons moving on the open field lines in the LC zone writes as \cite{OR2,OR1}
\begin{equation}
\label{gcor} 
\gamma_{cor}\simeq 1.5\times 10^7\times\left(\frac{\gamma_0}{10^4}\times\frac{sin^3\alpha}{\\cos\alpha}\right)^{1/2}\times\left(\frac{P}{P_{_{Cr}}}\right)^{5/4}\times\left(\frac{\dot{P}}{\dot{P}_{_{Cr}}}\right)^{1/4},
\end{equation} 
where the initial Lorentz factor is normalized by its typical value in the Crab-like pulsars \cite{OR1}, $P$ and $\dot{P}$ are normalized by the Crab pulsar's parameters, $P_{_{Cr}}\simeq 0.0332$ s and $\dot{P}_{_{Cr}}\simeq 4.21\times10^{-13}$ ss$^{-1}$. The condition $r\simeq r_{_L}/\sin\alpha$ has been taken into account  and we considered the open field lines with curvature radii exceeding the LC length scale, which means that in the framework of this approximation the field lines are assumed to be straight. As it is clear from the aforementioned expression in the Crab-like pulsar's magnetospheres the particles might achieve extremely high energies. One also can straightforwardly show that for normal pulsars ($P\simeq 1$ s, $\dot{P}\simeq 10^{-15}$ ss$^{-1}$) the co-rotation condition might guarantee the maximum attainable relativistic factor of the order of $2.6\times 10^5$.

\subsubsection{Emission constraints}

The approach presented in the previous section is simplified for several reasons: in realistic astrophysical scenarios dynamics of particles is strongly affected either by the IC scattering of relativistic particles against soft photons or curvature radiation. Unlike them, the synchrotron mechanism can not impose constraints on the motion of electrons. In particular, by taking the single particle synchrotron power into account, $P_{syn}\simeq2e^2\omega_B^2\gamma^2/3c$, where $\omega_B = eB/mc$ is the cyclotron frequency, the synchrotron cooling time-scale, $t_{syn}\simeq\gamma mc^2/P_{syn}$ is of the order of $6\times 10^{-9}$ sec for $\gamma =10^5$. This value is very small compared to the kinematic time-scale of pulsars (rotation period), therefore immediately after becoming relativistic the electrons lose their perpendicular momentum, go to the ground Landau level and continue sliding along the field lines, which in turn guarantees the centrifugal mechanism of acceleration.

In the previous subsection we have shown that efficiency of rotation energy pumping significantly increases in the LC area. If one assumes a straight co-rotating field line inclined by the angle $\alpha$ with respect to the rotation axis, then the acceleration time-scale on the LC
\begin{equation}
\label{tac1} 
t_{acc} = \frac{\gamma}{d\gamma/dt},
\end{equation} 
reduces to
\begin{equation}
\label{tac2} 
t_{acc}\simeq\frac{r_{_L}}{2c\sin\alpha}\times\left(1-\frac{r_0^2\sin^2\alpha}{r_{_L}^2}\right)^{1/2}\times\left(\frac{\gamma_0}{\gamma}\right)^{1/2}.
\end{equation} 
From Eq. (\ref{gcor}) it is evident that in the nearby zone of the LC the relativistic factor increases. This in turn, leads to the decrease of the acceleration time-scale (see Eq. (\ref{tac2})), indicating high efficiency of the acceleration process. On the other hand, the pulsar magnetosphere is full of soft - thermal photons originating from the neutron star's surface and the IC scattering seems to be important. In this context one has to note that very often the role of IC mechanism with thermal photons is supposed to be negligible because it has been accepted that the photons and relativistic electrons travel in the same - radial direction \cite{morini}. Contrary to this, following the idea, originally proposed by Gold (see \cite{gold1,gold2}), strong magnetic field (see the previous subsection) guarantees the frozen-in condition of electrons almost up to the LC zone resulting in the co-rotation of particles, having only the azimuthal velocity component, therefore, the angle of interaction is $\sim 90^0$ and consequently the IC scattering of electrons against thermal photons should be significant.

In the standard theory, the temperature of neutron star's surface strongly depends on their age, $\tau = P/(2\dot{P})$ \cite{ZH}
$$T(\tau)\simeq 5.9\times 10^5\times\left(\frac{10^6\; yr}{\tau}\right)^{0.1} K \;\;\;\; for \; \tau\leq 10^{5.2} \;yr,
$$
\begin{equation}
\label{temp} 
T(\tau)\simeq 2.8\times 10^5\times\left(\frac{10^6\; yr}{\tau}\right)^{0.5} K \;\;\;\; for \; \tau> 10^{5.2} \;yr,
\end{equation} 
therefore, for the Crab-like pulsars the age is of the order of $10^3$ yr and the surface temperature is $T\simeq 1.2\times 10^6$ K and for the normal pulsars ($P = 1$ s, $\dot{P} = 10^{-15}$ ss$^{-1}$) the age is three orders of magnitude bigger and consequently $T\simeq 2.4\times 10^5$ K. For the peak frequencies $\nu\simeq 2.8\;kT/h$ and the corresponding photon energies one has $\nu \simeq 6.7\times 10^{16}$ K and $\epsilon_{max}\simeq 0.3$ KeV for the Crab-type pulsars and $\nu \simeq 4.1\times 10^{15}$ K and $\epsilon_{max}\simeq 0.02$ keV for the normal pulsars respectively (here $h$ denotes the Planck's constant and $k$ is the Boltzmann's constant). In general, the IC scattering might occur in two extreme regimes, in the so-called Thomson regime for very low energy particles and in the Klein-Nishina regime, for very energetic electrons. In particular, the former takes place if the following condition is satisfied $\gamma\epsilon_{ph}/(mc^2) << 1$ \cite{blum} and in the latter case $\gamma\epsilon_{ph}/(mc^2) >> 1$, where $\epsilon_{ph}$ is the photon's energy. Then, one can straightforwardly check that for both class of pulsars, where we have already shown that $\gamma \simeq 10^7$ (Crab-like pulsars) and $\gamma \simeq 10^5$ (normal pulsars) the IC process occurs in the KN regime when the photons gain almost the total energy of electrons. Then, after taking into account the spherically symmetric photon field, the KN Compton power is given by \cite{blum}
\begin{equation}
\label{KN} 
P_{_{KN}}\simeq\frac{\sigma_{_T}\left(mckT\right)^2}{16\hbar^3}\left(\ln\frac{4\gamma kT}{mc^2}-1.981\right)\left(\frac{r_s}{r}\right)^2.
\end{equation} 
Logarithmic dependence on the Lorentz factor implies that the IC cooling time-scale in the LC area
\begin{equation}
\label{tic} 
t_{_{IC}} = \frac{\gamma mc^2}{P_{_{KN}}}\propto\gamma
\end{equation} 
is a continuously increasing function of $\gamma$, which indicates that the IC process does not impose any significant constraints on the maximum attainable energies. Therefore, the maximum energy is limited by the co-rotation constraint and consequently after scattering photons will have energies 
\begin{equation}
\label{energy} 
\epsilon_{_{ph}}\simeq 7.6 \times\left(\frac{\gamma_0}{10^4}\times\frac{sin^3\alpha}{\\cos\alpha}\right)^{1/2}\times\left(\frac{P}{P_{_{Cr}}}\right)^{5/4}\times\left(\frac{\dot{P}}{\dot{P}_{_{Cr}}}\right)^{1/4},\; TeV
\end{equation} 
in Crab-type pulsars and $130$MeV in normal pulsars. 

Despite the fact that the IC mechanism does not affect the maximum achievable energy of electrons, the produced VHE luminosity might be quite high. For calculating this particular quantity, one should emphasise that the acceleration is extremely efficient in a narrow thickness, where the relativistic factor asymptotically increases and the corresponding scaling factor - rate of change, $d\gamma/dr$, increases as well. Therefore, acceleration is more efficient for highly nonuniform $\gamma$'s and hence the corresponding thickness length-scale can be estimated as to be $d\simeq \frac{\gamma}{d\gamma/dr}\simeq \gamma_0 r_{_L}/(2\gamma)$ where we have taken into account Eq. (\ref{gcor}) for $r_0 = 0$. The azimuthal length-scale involved in the emission process is of the order of $\delta l\simeq r_{_L}\theta$, where $\theta\simeq 2\pi\xi/10$ ($\xi<10$) and consequently for the active volume one has $\Delta V\simeq (\delta l)^2d$.  

We have discussed in the previous subsection that the pair cascading is very efficient. The corresponding number density of electron-positron pairs is significantly increased and as is shown in \cite{GJ} the multiplicity factor equals $1/(1-r^2/r_{_L}^2)$, which combined with Eq. (\ref{gamma}) leads to $M = \gamma/\gamma_0$. After taking into account all these quantities the expression for the bolometric IC luminosity is given by
$$L_{_{IC}}\simeq 2n_{_{GJ}}M\Delta VP_{_{KN}}\simeq $$
\begin{equation}
\label{LIC} 
\simeq 6.1\times 10^{29}\times\xi^2\times\sin^5\alpha\times\left(\frac{T}{1.2\times 10^6 \;K}\right)^2\times\left(\frac{P_{_{Cr}}}{P}\right)^{5/2}\times\left(\frac{\dot{P}}{\dot{P}_{_{Cr}}}\right)^{1/2}\; ergs\;s^{-1},
\end{equation} 
where we have taken into account that in generation of the TeV emission the most energetic photons are involved. The normal pulsars (with $T\simeq 2.4\times 10^5$ K) in the VHE domain ($\sim 100$ MeV) might provide luminosity of the order of $\sim 1.6\times 10^{21}$ ergs s$^{-1}$.

Even if the field lines are almost straight, due to the co-rotation condition the particles follow the field lines and therefore in the laboratory frame of reference their trajectories might be significantly curved leading to the mechanism of curvature radiation. In the context of pulsars this mechanism has been investigated in \cite{gang1,gang2} where the authors discussed in detail the generation of radio emission by means of the curvature radiation mechanism. We will see that this process might account for the generation of much higher energies.

The energy loss rate by means of the curvature radiation of the electron is given by
\begin{equation}
\label{Pcur} 
P_c = \frac{2}{3}\frac{e^2c}{R_c^2}\gamma^4,
\end{equation} 
where $R_c$ denotes the radius of the curvature. The characteristic cooling time-scale, $t_{c}\equiv \gamma mc^2/P_c$, then becomes
\begin{equation}
\label{tcur} 
t_c = \frac{3mcR_c^2}{2e^2\gamma^3}. 
\end{equation} 
It is clear from this expression that $t_c$ is vanishing faster than the acceleration time-scale, $t_{acc}\propto 1/\gamma^{1/2}$, (see Eq. (\ref{tac1})), therefore, the maximum attainable Lorentz factor, $\gamma_c$, is achieved when $t_{acc}\simeq t_c$, leading to the following expression of $\gamma_c$
$$\gamma_c\simeq\frac{1}{\gamma_0^{1/5}}\times\left(\frac{3mPc^3\sin\alpha}{2\pi e^2}\right)^{2/5}\times\left(\frac{R_c}{r_{_L}}\right)^{4/5}\simeq$$
\begin{equation}
\label{gcur} 
\simeq 4.9\times 10^7\times\left(\frac{10^4}{\gamma_0}\right)^{1/5}\times\left(\frac{R_c}{r_{_L}}\right)^{4/5}\times\left(\frac{P}{P_{_{Cr}}}\right)^{2/5}\times\sin^{2/5}\alpha,
\end{equation} 
where we have taken into account that the curvature radius, $R_c$ of the trajectory in the nearby zone of the LC is of the order of $r_{_L}$, which is clearly seen from Eq. (\ref{gamma}). In particular, close to the mentioned area the Lorentz factor asymptotically increases and the linear velocity of rotation tends to the speed of light, therefore, the radial velocity tends to zero and consequently the curvature radius equals the LC radius. 

Since the derived value of the maximum relativistic factor is higher than the Lorentz factor limited by the co-rotation constraint (see Eq. (\ref{gcor})), the latter is responsible for reaching the maximum energy of electrons, leading to the following energy of curvature photons \cite{OR1} 
$$\epsilon_{cur}\simeq\frac{3ch}{4\pi R_c}\gamma^3\simeq$$
\begin{equation}
\label{ecur} 
\simeq 0.6\times\left(\frac{\gamma_0}{10^4}\times\frac{sin^3\alpha}{\\cos\alpha}\right)^{3/2}\times\left(\frac{P}{P_{_{Cr}}}\right)^{11/4}\times\left(\frac{\dot{P}}{\dot{P}_{_{Cr}}}\right)^{3/4}\; GeV.
\end{equation} 
As we see, the curvature radiation provides the emission in the GeV range for the millisecond pulsars. For the total curvature luminosity one obtains
$$L_c\simeq 2n_{_{GJ}}M\Delta VP_c\simeq $$
\begin{equation}
\label{Lcur} 
\simeq 2.8\times 10^{34}\times\left(\xi\times\frac{r_{_L}}{R_c}\times\frac{\gamma_0}{10^4}\times\frac{\sin^3\alpha}{\cos\alpha}\right)^2\times\left(\frac{P}{P_{_{Cr}}}\right)^{5/2}\times\left(\frac{\dot{P}}{\dot{P}_{_{Cr}}}\right)^{3/2} \; ergs\;s^{-1}.
\end{equation} 
Despite the fact that the curvature emission mechanism is not responsible for limiting the maximum attainable energy of particles, most of the energy is emitted by means of it.

For the normal pulsars ($P\simeq 1$ s, $\dot{P}\simeq 10^{-15}$ ss$^{-1}$) one can straightforwardly check that the co-rotation still remains the major mechanism limiting the maximum energy and the curvature radiation provides emission in the upper UV domain ($\epsilon_c\simeq 100$ eV) with the luminosity of the order of $2.4\times 10^{22}$ ergs s$^{-1}$.

\subsection{Parametric excitation of electrostatic waves and Langmuir Landau centrifugal drive}

In the previous subsection we have shown that centrifugal mechanism of particle acceleration might be very efficient in millisecond as well as normal pulsar's magnetospheres. It is worth noting that different species of centrifugally accelerated particles experience different forces leading to charge separation and a consequent generation of the electrostatic field. On the other hand, as it is found in \cite{incr,group} the relativistic centrifugal force is time dependent and therefore, the excitation process is parametrically amplified. 

For studying dynamics of a plasma flow following the magnetic field lines, it is assumed that the field lines are almost rectilinear, which is satisfied on distances less than the curvature radius. For simplicity one can consider dynamics of plasma particles in the local frame of reference of a rotating pulsar. 
If one considers the $1+1$ formalism \cite{paradigm} in the framework of a single particle approximation, for the chosen reference frame the zero angular momentum observers - ZAMOs, will measure the following proper time $d\tau\equiv\zeta dt$, where $dt$ is the universal time, $\zeta\equiv\sqrt{1-\Omega_{ef}^2r^2/c^2}$ is the so-called lapse function. The equation of motion then is given by the following equation \cite{incr}
\begin{equation}
\label{eqmo1} 
\frac{d{\bf p}}{d\tau} = \gamma{\bf g}+\frac{e}{m}\left({\bf E}+\frac{1}{c^2}{\bf v}\times{\bf B}\right),
\end{equation} 
where ${\bf E}$ is the electric field, ${\bf p}\rightarrow{\bf p}/mc$ is the dimensionless momentum, $\gamma{\bf g}$ is the effective centrifugal force \cite{chedia} with  ${\bf g} = - {\bf\nabla}\zeta/\zeta$ and in the relativistic factor $\gamma = \left(1-V^2/c^2\right)^{1/2}$ the velocity is defined according to the $1+1$ approach ${\bf V} = d{\bf r}/d\tau$. 
In order to rewrite the aforementioned equation for a magnetospheric fluid one should emphasise a complex composition of plasmas. In Sec. 2.1.1. we have briefly discussed that initially accelerated particles radiate by means of the curvature process. Newly produced photons under certain conditions might generate secondary electron positron pairs. Finally magnetospheric $e^+, e^-$ population may be divided into two major components: (I) the bulk component - the basic plasma mass with relatively mild Lorentz factors and (II) the beam component with high values of relativistic factors.Then, if one takes into account the identity $d/d\tau = \partial/(\zeta\partial t)+\left({\bf V}{\bf\nabla}\right)$, Eq. (\ref{eqmo1}) will reduce to 
\begin{equation}
\label{eqmo2} 
\frac{1}{\zeta_i}\frac{\partial{\bf p_i}}{\partial t}+\left({\bf V_i}{\bf\nabla}\right){\bf p_i}= \gamma_i{\bf g_i}+\frac{e_i}{m}\left({\bf E}+\frac{1}{c}{\bf V_i}\times{\bf B}\right),
\end{equation} 
for several species of particles denoted by $i = b, e, p$, where $b$ corresponds to bulk particles (mostly electrons), $e$ - to electrons and $p$ - to positrons. One should note that ZAMO's momentum is coincident with the momentum measured in the inertial frame of reference. Indeed, by combining the definition of momentum, ${\bf p} = \gamma {\bf V}$, where $\gamma = \zeta\gamma '$ and ${\bf V'} = d{\bf r}/d\tau$ (prime denotes the quantities in the inertial frame of reference) one can straightforwardly show ${\bf p} = {\bf p'}$. Then for the Eq. (\ref{eqmo2}) rewritten in the inertial frame one has (omitting primes)
\begin{equation}
\label{moment} 
\frac{\partial{\bf p_i}}{\partial t}+\left({\bf v_i}{\bf\nabla}\right){\bf p_i}= - c^2\zeta_i\gamma_i{\bf\nabla}\zeta_i+\frac{e_i}{m}\left({\bf E}+\frac{1}{c}{\bf v_i}\times{\bf B}\right).
\end{equation} 
As we have already discussed, the centrifugal force, acting on different species, leads to charge separation, which automatically creates the electrostatic field. Therefore, for describing the physical system momentum equation should be complemented by the continuity equation and the Poisson equation
\begin{equation}
\label{cont} 
\frac{\partial n_i}{\partial t}+{\bf\nabla}\left(n_i{\bf v_i}\right) = 0,
\end{equation} 
\begin{equation}
\label{pois} 
{\bf\nabla}{\bf E} = 4\pi\;\sum_{i}e_in_i,
\end{equation} 
where $n_i$ denotes the number density of a corresponding component. It is worth noting that strong magnetic field guarantees the frozen-in condition of plasmas, ${\bf E_0}+\frac{1}{c}{\bf v_{0i}}\times{\bf B_0} = 0$, which is supposed to be a leading state of the system. By taking into account this condition the momentum equation reduces to Eq. (\ref{dv}) which in a dimensional form is given by
\begin{equation}
\label{dv1} 
\frac{d^2r}{dt^2} = \frac{\Omega_{ef}^2r}{1-\frac{\Omega_{ef}^2r^2}{c^2}}\left[1-\frac{\Omega_{ef}^2r^2}{c^2}-2\frac{\upsilon^2}{c^2}\right],
\end{equation} 
where $\upsilon = dr/dt$ denotes the radial velocity. In \cite{MR} it has been shown that for ultrarelativistic particles ($\gamma>>1$) Eq. (\ref{dv1}) has the following solution
\begin{equation}
\label{r} 
r(t)\simeq\frac{c}{\Omega_{ef}}\sin\left(\Omega_{ef}t+\phi\right),
\end{equation} 
\begin{equation}
\label{v} 
\upsilon_0(t)\simeq c\cos\left(\Omega_{ef}t+\phi\right),
\end{equation} 
where $\phi$ denotes the phase. After taking into account these solutions, one can straightforwardly show that the analog of the centrifugal force, $- \zeta_i\gamma_i{\bf\nabla}\zeta_i$, is time dependent and drives the parametric excitation of the Langmuir waves. In this paper we intend to study the linear generation of the electrostatic waves. Therefore, all physical quantities are expanded around the leading state $\Psi^0$
\begin{equation}
\label{psi} 
\Psi\approx\Psi^0+\Psi^1,
\end{equation} 
where $\Psi = (n, {\bf v}, {\bf p}, {\bf E}, {\bf B})$ and $\Psi^1$ denotes a small perturbation of the corresponding quantity. It is straightforward to show that the anzatz
\begin{equation}
\label{anzatz1} 
\Psi^1(t,{\bf r}) = \Psi^1(t)\; exp\left(i{\bf kr}\right),
\end{equation} 
leads to the following linearized versions of Eqs. (\ref{moment}-\ref{pois}) 
\begin{equation}
\label{moment1} 
\frac{\partial p_i^1}{\partial t}+ ik\upsilon_0p_i^1 = \upsilon_0\Omega_{ef}^2rp_i^1+\frac{e_i}{m}E^1.
\end{equation} 
\begin{equation}
\label{cont1} 
\frac{\partial n_i^1}{\partial t}+ik\upsilon_{i0}n_i^1 + ikn_{i0}\upsilon_i^1 = 0,
\end{equation} 
\begin{equation}
\label{pois1} 
ik E^1 = 4\pi\;\sum_{i}e_in_i^1.
\end{equation} 
By introducing the so-called plasma quantities for perturbations $n_{pl}^1 = n_{e}^1-n_{p}^1$, $\upsilon^1_{pl} = {\upsilon^1_{e}}-{\upsilon^1_{p}}$ (the plasma component) with the unperturbed quantities of electrons and positrons $n^0_e = n^0_p$ and $\upsilon^0_{e} = \upsilon^0_{p}$, and considering the bulk component and the plasma component respectively, after straightforward but tedious mathematical manipulations one arrives at the coupled non-autonomous differential equations governing the centrifugal generation of the electrostatic instability \cite{MMOC}
\begin{equation}
\label{eq1} 
\frac{d^2N_1}{dt^2}+\omega_1^2N_1 = -\omega_1^2N_2 e^{i\Xi},
\end{equation} 
\begin{equation}
\label{eq2} 
\frac{d^2N_2}{dt^2}+\omega_2^2N_2 = -\omega_2^2N_1 e^{-i\Xi},
\end{equation} 
where $\omega_{1,2} = \sqrt{8\pi e^2n_{1,2}/m\gamma_{1,2}^3}$ and $N_1$ and $N_2$ relate to the first order perturbations of number densities of two species in the following way
\begin{equation}
\label{anzatz2} 
N_{\beta} = n_{\beta}\exp\left[\frac{ick}{\Omega}\sin\left(\Omega t+\phi_{\beta}\right)\right], \; \beta = 1,2,
\end{equation} 
where $\phi_{\beta}$ denote the corresponding phases and $\Xi$ is expressed by the following way
\begin{equation}
\label{xi} 
\Xi = b\cos\left(\Omega t+\phi_{+}\right),
\end{equation} 
and 
\begin{equation}
\label{b} 
b = 2\frac{ck}{\Omega}\sin\phi_-,
\end{equation} 
with $2\phi_{\pm} = \phi_1\pm\phi_2$. We have assumed that energy is almost uniformly distributed among the species, $n_1\gamma_1\simeq n_2\gamma_2\simeq n_{_{GJ}}\gamma_b$.
After Fourier transforming Eqs. (\ref{eq1},\ref{eq2}) by applying the tools developed in \cite{incr}, one can arrive to the following dispersion relation
\begin{equation}
\label{disp} 
-\omega^2+\omega_1^2+\omega_2^2\;J^2_0(b) =-\omega^2\omega_2^2\sum_{\mu\neq 0}\frac{J^2_{\mu}(b)}{\left(\omega+\mu\Omega\right)^2}.
\end{equation} 
From the obtained expression it is clear that the resonance frequency is given by $\omega_r = \left(\omega_1^2+\omega_2^2\;J^2_0(b)\right)^{1/2}$. Following the method described in \cite{incr}, expanding the frequency, $\omega = \omega_r+\Delta$, and emphasising that the near the resonance the basic contribution to the r.h.s. of the equation comes form the term $\omega_r = -\mu_0\Omega$, one obtaines the following equation
\begin{equation}
\label{incr1} 
\Delta^3 = \frac{\omega_r\;\omega_2^2}{2}\;J^2_{\mu_0}(b).
\end{equation} 
Solving this cubic equation, one can show that the imaginary part of the solution, thus the growth rate of the instability, writes as \cite{OMMC}
\begin{equation}
\label{incr2} 
\Gamma = \frac{\sqrt{3}}{2} \left(\frac{\omega_r\;\omega_2^2}{2}\right)^{1/3}\;J^{2/3}_{\mu_0}(b).
\end{equation} 
Already amplified electrostatic waves will inevitably Landau damp with the following increment \cite{volok}
\begin{equation}
\label{incr3} 
\Gamma_{LD} = \frac{n_{_{GJ}}\gamma_b\; \omega_b}{n_p\gamma_p^{5/2}}.
\end{equation} 
It is worth noting that the most optimal regime is when both growth rates are of the same order of magnitude. For example, if one considers the Crab pulsar with the possible maximum relativistic factor provided by the direct CA, $1.5\times 10^7$, (see Eq. (\ref{gcor})), one can see that two streams with $\gamma_1\simeq10^4$
and  $\gamma_2\simeq 7\times10^4$ drive the electrostatic instability with the corresponding time-scale, $1/\Gamma$, of the order of $0.01$sec. For the aforementioned stream parameters (considering the stream 1 component as to be the plasma component) both growth rates are equal and since the time-scale is less than the kinematic time-scale (period of rotation) of the Crab pulsar, the process is efficient. Therefore the energy pumped by the primary beam by means of the Landau process should be enormous.

In particular, by considering the force responsible for the particle acceleration, $F_r\simeq 2mc\Omega\;\zeta^{-3}$ \cite{OMMC}, the energy accumulated in the primary beam after Landau damping is given by \cite{OMMC}
\begin{equation}
\label{emax} 
\epsilon\simeq\frac{n_pF_r\delta r}{n_{_{GJ}}}
\end{equation} 
and for the aforementioned two streams is of the order of $3.7$PeV. Here $\delta r\simeq c/\Gamma$ is the corresponding length-scale where the process of re-acceleration takes place. After applying the same mechanism to normal pulsars ($P\simeq 1$ s, $\dot{P}\simeq 10^{-15}$ ss$^{-1}$) with the beam Lorentz factor of the order of $2.6\times 10^5$, one can straightforwardly check that two streams with $\gamma_1\simeq10^2$
and  $\gamma_2\simeq 5\times10^2$, lead to the parametric instability of generation of Langmuir waves with a high growth rate $\Gamma\sim\Gamma_L\simeq 46\;s^{-1}$. Consequently, by Eq. (\ref{emax}) one can check that the re-acceleration process in normal pulsars, although cannot guarantee such high energies as in the Crab-like pulsars, but still energies will be in the VHE domain, $360$ GeV. 

In subsection 2.1.2. we have shown that the IC scattering and synchrotron mechanism can not affect the dynamics of particles. The Langmuir waves propagate along the magnetic field lines, and the resulting re-acceleration will not cause the pitch angle significantly suppressing the synchrotron process. We have already discussed that from observations it is evident that cosmic particles leave the pulsar's magnetospheres, which indicates that they are in the force-free regime. Therefore, as is studied in \cite{jet,ODM,bog} the field lines lag behind the rotation, having the shape of the Archimedean spiral characterised by the rectilinear trajectories in the laboratory frame of reference. This means that the curvature emission is also strongly suppressed and does not impose any significant constraints on maximum achievable energies of electrons.

\section{Conclusions}
Centrifugal processes occurring in the pulsar's magnetospheres might lead to new and very interesting phenomena. This may allow new insights into the physics of these extreme environments. The detection of deceleration of pulsars indicates that a significant fraction of neutron stars' rotation energy is extracted from the central object. A certain fraction of it can be pumped into the magnetospheic electrons and electromagnetic radiation.

In particular, we have demonstrated that direct CA is a very efficient mechanism providing the observed jet velocity of the Crab pulsar. this process is responsible also for generation of relativistic particles with the Lorentz factors of the order of $10^5$ (for normal pulsars) and $10^7$ (for millisecond Crab-like pulsars).

It has been found that IC scattering might provide photon energies of the order from $100$ MeV up to several TeV depending on a pulsar's type. The total luminosity for Crab-like pulsars might be at least $10^{29}$ ergs s$^{-1}$ and for relatively slowly rotating pulsars - $10^{21}$ ergs s$^{-1}$.

Curvature mechanism, although provides electromagnetic radiation in the GeV (Crab-like pulsars) and UV (normal pulsars) domain, the corresponding luminosities are higher than in the IC scattering: $10^{34}$ ergs s$^{-1}$ (Crab-like pulsars) and $10^{22}$ ergs s$^{-1}$ (Normal pulsars).

We have also examined the effects of CA on the excitation of Langmuir waves. It has been found that by means of the strong centrifugal force in the LC zone the electrostatic waves are parametrically excited. The corresponding instability time-scale is much less than the rotation period of a neutron star, indicating high efficiency of the driving process.

The amplification of the Langmuir waves will be terminated by means of the Landau damping resulting in extremely high energies of particles. We have found that the electrons in Crab-like pulsars after re-acceleration via the mechanism of Langmuir-Landau-centrifugal-drive might reach energies in the PeV domain. For the normal pulsars the achieved energies of electrons is of the order of several hundreds of GeV.

The centrifugal parametric excitation of Langmuir waves has a very interesting consequence. In particular, as we have shown, the growth rate of the instability is so large that the electrostatic waves are efficiently amplified. During this process the field can reach the Schwinger limit and consequently an efficient pair creation will take place. This in turn might significantly influence physical processes in the $\gamma$-ray pulsar's magnetospheres and sooner or later we are going to study this particular problem.


\end{document}